\documentclass{aa}
\usepackage[varg]{txfonts}
\usepackage{graphicx}
\usepackage{natbib}
\usepackage{epstopdf}
\usepackage{epsfig}
\bibpunct{(}{)}{;}{a}{}{,} 

\begin{document}

\title{Inconsistency of the  Wolf sunspot number series around 1848}

\author{Raisa Leussu\inst{1,2}
\and Ilya G. Usoskin\inst{1,2}
\and Rainer Arlt\inst{3}
\and Kalevi Mursula\inst{1}}

\institute{Department of Physics, University of Oulu, Finland
\and Sodankyl\"a Geophysical Observatory (Oulu unit), University of Oulu, Finland
\and Leibniz Institute for Astrophysics Potsdam, Germany}

\date{Received $<$date$>$ /
Accepted $<$date$>$}

\abstract {}
{Sunspot number is a benchmark series in many studies, but may still contain inhomogeneities and inconsistencies.
 In particular, an essential discrepancy exists between the two main sunspot number series,
  Wolf (WSN) and group (GSN) sunspot numbers, before 1848.
 The source of this discrepancy has so far remained unresolved.
 However, the recently digitized series of solar observations in 1825--1867 by Samuel Heinrich Schwabe, who was the primary observer of the WSN before 1848, makes such an assessment possible.}
{We construct sunspot series, similar to WSN and GSN, but using only Schwabe's data.
 These series, called WSN-S and GSN-S, respectively, were compared with the original WSN and GSN
  series for the period 1835--1867 to look for possible inhomogeneities.}
{We show that: (1) The GSN series is homogeneous and consistent with the Schwabe data throughout the entire studied period;
(2) The WSN series decreases by roughly $\approx 20$\% around 1848 caused by the change of the
 primary observer from Schwabe to Wolf and an inappropriate individual correction factor used for Schwabe in the WSN;
 (3) This implies a major inhomogeneity in the WSN, which needs to be corrected by reducing its values by 20\% before 1848;
 (4) The corrected WSN series is in good agreement with the GSN series.
This study supports the earlier conclusions that the GSN series is
 more consistent and homogeneous in the earlier part than the WSN series.} {}

\maketitle

\section{Introduction}

Sunspot numbers form the longest series of direct astronomical observations.
Started already in 1610, systematic counts and drawings of sunspots have been carried out
 through centuries by an army of professionals and amateurs \citep[for reviews, see,][]{hathawayLR, usoskin13}.
The tremendous work by Rudolf Wolf of Z\"urich in the second half of the 19th century resulted in
 creation of the first official sunspot number series, called the Wolf sunspot numbers (WSN).
The core of the WSN series is based on the observations made by Rudolf Wolf and his successors after 1848 and
 continued to the present times, making the WSN series continuous since 1848.
However, the quality of the WSN series before 1848 is sometimes doubted as it is a result of
 compilation, by Wolf, of different observations and records and includes poorly grounded
 re-calibration \citep[e.g.][]{sonett83,wilson88,wilson98,usoskin13}.
Another series of sunspot activity, called the group sunspot numbers (GSN) was compiled much more recently \citep{hoyt98}.
It is methodologically different from the WSN and contains a much greater amount of original information
 than the WSN, and is often considered to be more homogeneous and representative in the early part of the series than the WSN \citep[e.g.,][]{hathaway02,usoskin13}.

Since the sunspot numbers form a benchmark series for many research topics and practical applications, from statistical analysis to solar and terrestrial physics, it is crucially important to verify the homogeneity of the sunspot number series, particularly on the long time scale.
However, until now it has been difficult to directly test the quality of the two sunspot number series or their relative correctness in the earlier part \citep[e.g.,][]{wilson98}.

Thanks to the recent work by \citet{arlt13}, all individual drawings of one of the most
 famous and scrupulous solar observers Samuel Heinrich Schwabe, in 1825--1867 have now been
 digitized, forming an independent homogeneous series of sunspot observations.
Accordingly, we use this new data based on Schwabe's observations to study the consistency of
 the WSN and GSN series for the period between 1835 and 1867.

\section{Sunspot number series}
\label{Sec:SN}

\subsection{Wolf sunspot numbers WSN}

\begin{figure}
\centering \resizebox{8cm}{!}{\includegraphics{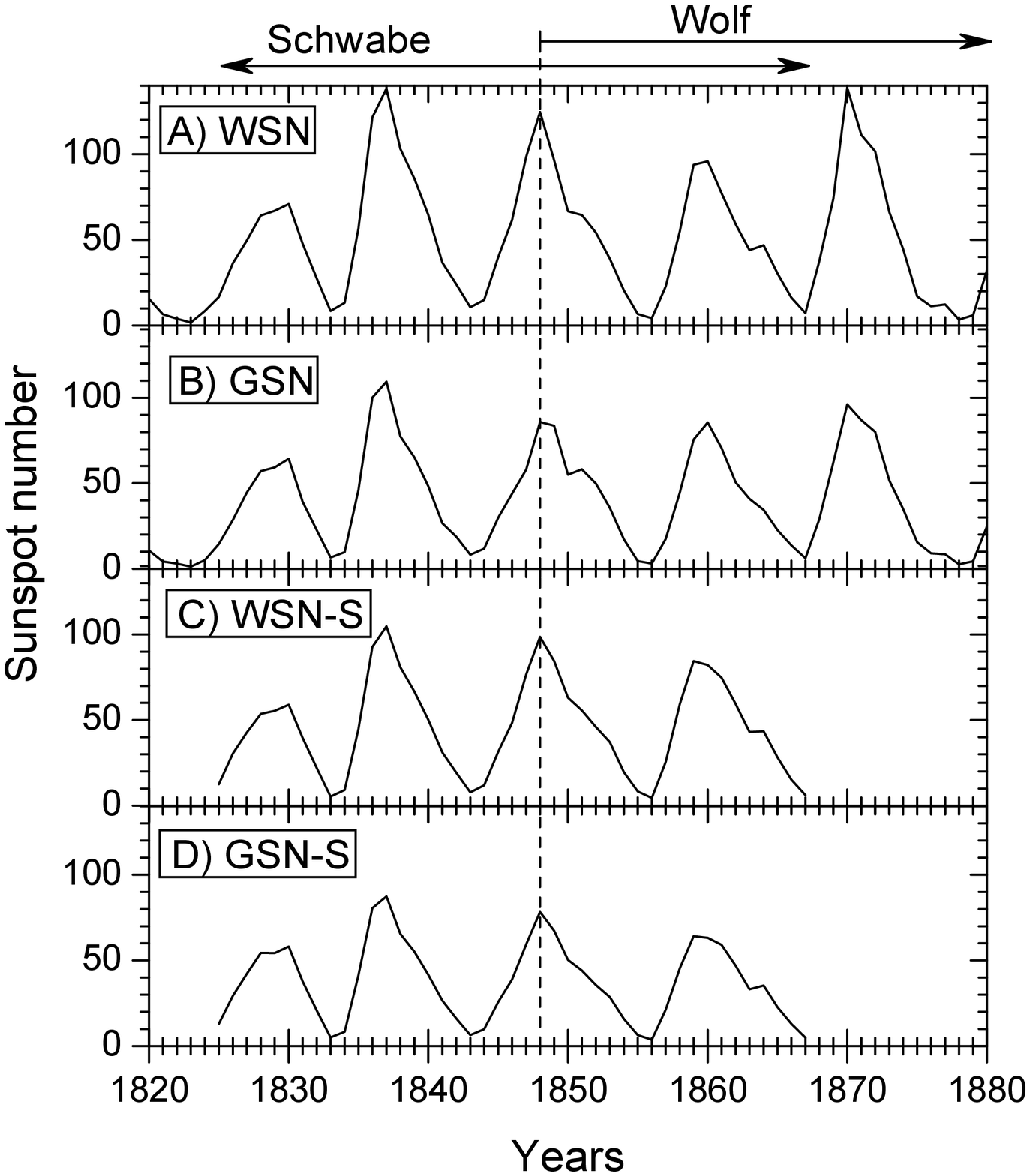} }
\caption{Annual sunspot numbers: a) the official Wolf sunspot number WSN; b) the Group sunspot number GSN; c) the Wolf sunspot number calculated from Schwabe data, WSN-S; d) the group sunspot number calculated from Schwabe data, GSN-S. Periods of sunspot observations by Schwabe and Wolf are indicated by arrows at the top. The year 1848 is also noted as vertical dashed line.}
\label{Fig:SN}
\end{figure}

The Wolf sunspot number, also called the  Z\"urich sunspot numbers series is calculated as
\begin{equation}
WSN = k(10\cdot G + N),
\label{Eq:WSN}
\end{equation}
where $G$ and $N$ are the number of sunspot groups and the number of individual sunspots reported
 by the selected observer for a given day respectively, and $k$ is the observer's correction factor
 accounting for his experience and the quality of his instrument.
The WSN uses only one observation per day, by the primary observer, selected by Wolf and his
 successors.
If the primary observer did not report solar observations for a particular day, secondary, tertiary
 etc., observers are used.
The hierarchical system of observers \citep{waldmeier61} was established with the purpose of homogeneity of the series.
This system of the WSN was changed in the late 20th century for the International sunspot number \citep{clette07}
 but was still used during the period under investigation.
It is important that Schwabe was the primary observer for the Wolf series for
 the period 1826 through 1847.
In 1848--1893 Wolf himself was the primary observer.
Thus, the period studied here is characterized by two primary observers - Schwabe before 1848
 and Wolf after that.
The WSN series for the period under question is shown in Figure~\ref{Fig:SN}a.

\subsection{Group sunspot numbers GSN}

The group sunspot number, introduced by \citet{hoyt98}, is based only on the number of sunspot groups,
 without individual spots.
The GSN is supposed to be more homogeneous in the long term than the WSN since the sunspot groups are more robustly defined than individual spots.
Another principal difference between the GSN and WSN is that the former uses all the available observations
 for a given day, not only the one by the primary observer.
The definition of the GSN is
\begin{equation}
GSN={12.08\over n}\sum_{i=1}^n{k_i\cdot G_i},
\label{Eq:GSN}
\end{equation}
where $G_i$ is the number of sunspot groups as reported by $i$-th observer, $k_i$ is the
 individual correction factor of the observer, and $n$ is the number of observers whose
 data are available for the given day. Factor 12.08 is included in order to normalize the GSN to the same absolute level as the WSN in 1874-1976 \citep{hoyt98}.
The GSN series is usually considered as more homogeneous in its earlier part than the WSN
 \citep{letfus99,hathaway02,usoskin_Rev_04}.
The GSN series around the period under question is shown in Figure~\ref{Fig:SN}b.

\subsection{Schwabe sunspot numbers WSN-S and GSN-S}
\label{Sec:S}

Samuel Heinrich Schwabe was an amateur astronomer, who performed daily observations of the Sun at his home in Dessau, Germany, nearly continuously from 1825 till 1868.
His sunspot observations contain over 8000 drawings of the solar disk and over 3000 verbal descriptions.
These data, including drawings, have been recently digitized and tabulated, as reported by \citet{arlt13}.
The data set produced from these measurements contain information on the date, heliospheric coordinates
 and subjective size of each individual spot drawn by Schwabe.
The daily numbers of groups and spots were calculated from these data to be compared with the official indices for this period of time.

In addition to drawings, Schwabe also defined sunspot groups for each day.
Rudolf Wolf, who used Schwabe as the primary observer from 1826 to 1847, listed the number of spots and
 groups for almost each day from Schwabe notes.
In order to estimate the relation between Schwabe's group definition and the one Rudolf Wolf presumably had,
  we compared Schwabe's groups with those defined by Rudolf Wolf during the same time period \citep{wolf1850}.
Out of all the 8401 days, when comparison is possible, there are only 19 cases when Schwabe had assigned one
 group more than Wolf, and 25 cases, when it was one group less.
Considering the small number (0.5\%) of differing assignments it is obvious that Wolf and
 Schwabe had a very similar perception on the definition of sunspot groups, and these assignments can be used in further calculations.
We note that their definition is somewhat different from the modern definition of sunspot groups, but this does not affect the results, since the Schwabe and Wolf numbers are compared with each other, not calibrated with today's values.

Using Schwabe's definition of the groups and individual spots, we
 computed analogs of the WSN and GSN series based solely on Schwabe's data.
The daily WSN-S and GSN-S were calculated by means of Eq.~\ref{Eq:WSN} and Eq.~\ref{Eq:GSN},
 respectively, using the daily numbers of spots and groups defined in Schwabe's data.
The correction factors $k$ were set to unity.
The monthly averages were acquired by calculating mean values
 of WSN-S and GSN-S from the daily data, and the annual averages by calculating mean values of the monthly data.
The resulting annual means of WSN-S and GSN-S are shown in Fig.~\ref{Fig:SN} c and d.

\begin{figure}
\centering \resizebox{8cm}{!}{\includegraphics{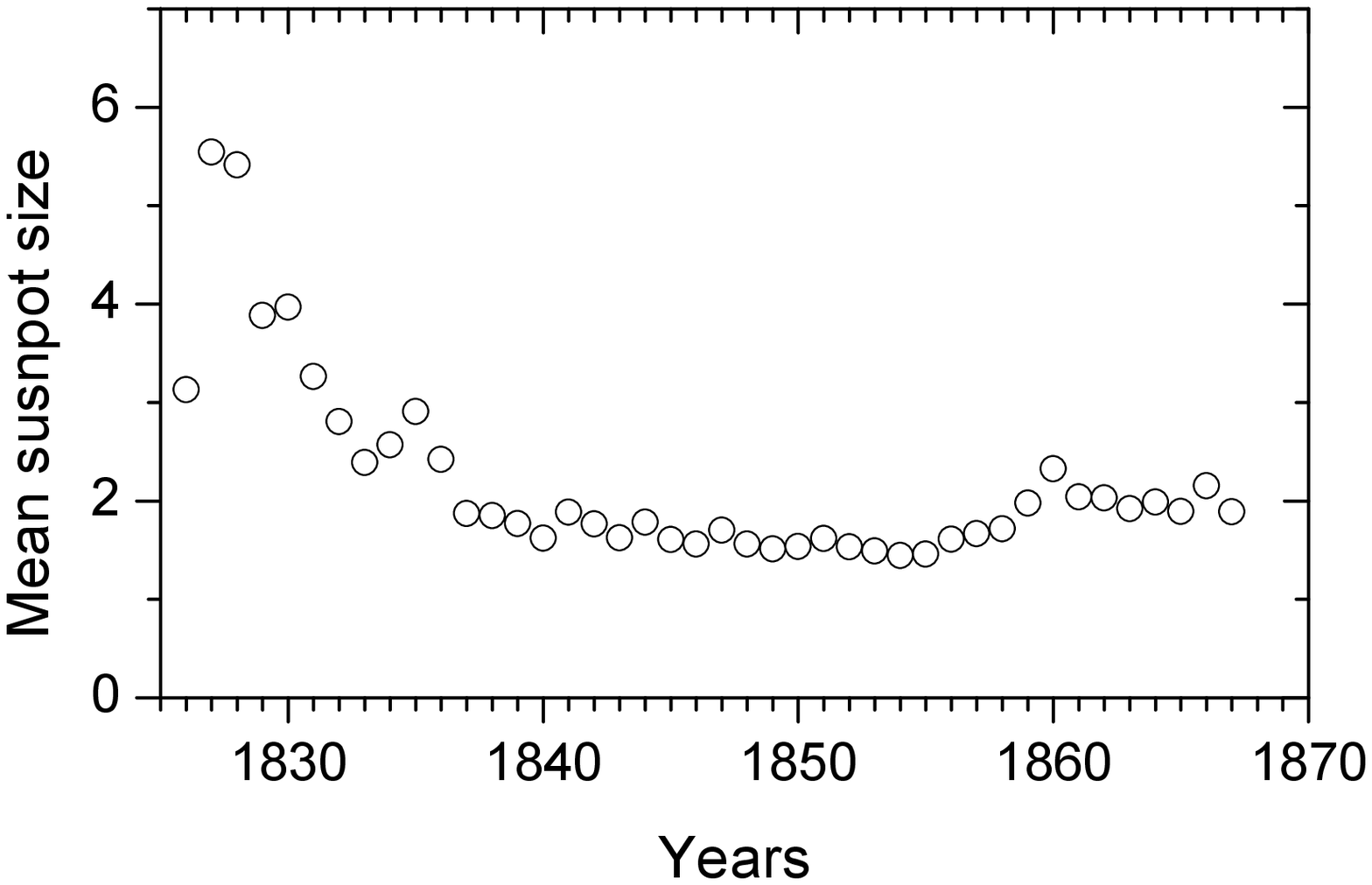} }
\caption{The mean size (in arbitrary units used during the digitization)
 of sunspots drawn by Schwabe.}
\label{Fig:size}
\end{figure}

We note that during the first ten years the accuracy of Schwabe's observations was gradually changing,
 most probably because of the changing drawing style.
This can be observed for example from Fig.~\ref{Fig:size} where the mean size of spots drawn
 by Schwabe is shown as function of time.
During 1826--1835 he mostly plotted large spots, while after ca. 1835 he recorded also spots of smaller size homogeneously.
This suggests that, for whatever reason, many small spots might have been left unnoticed before 1835.
This consequently reduces the WSN values compared to GSN at this time, since the number of sunspot groups can be identified
 more robustly.
Therefore, for the present analysis we will limit ourselves to the period of 1835--1867,
 when Schwabe's series is homogeneous.

\section{Comparing different sunspot numbers}

\begin{figure}
\centering \resizebox{8cm}{!}{\includegraphics{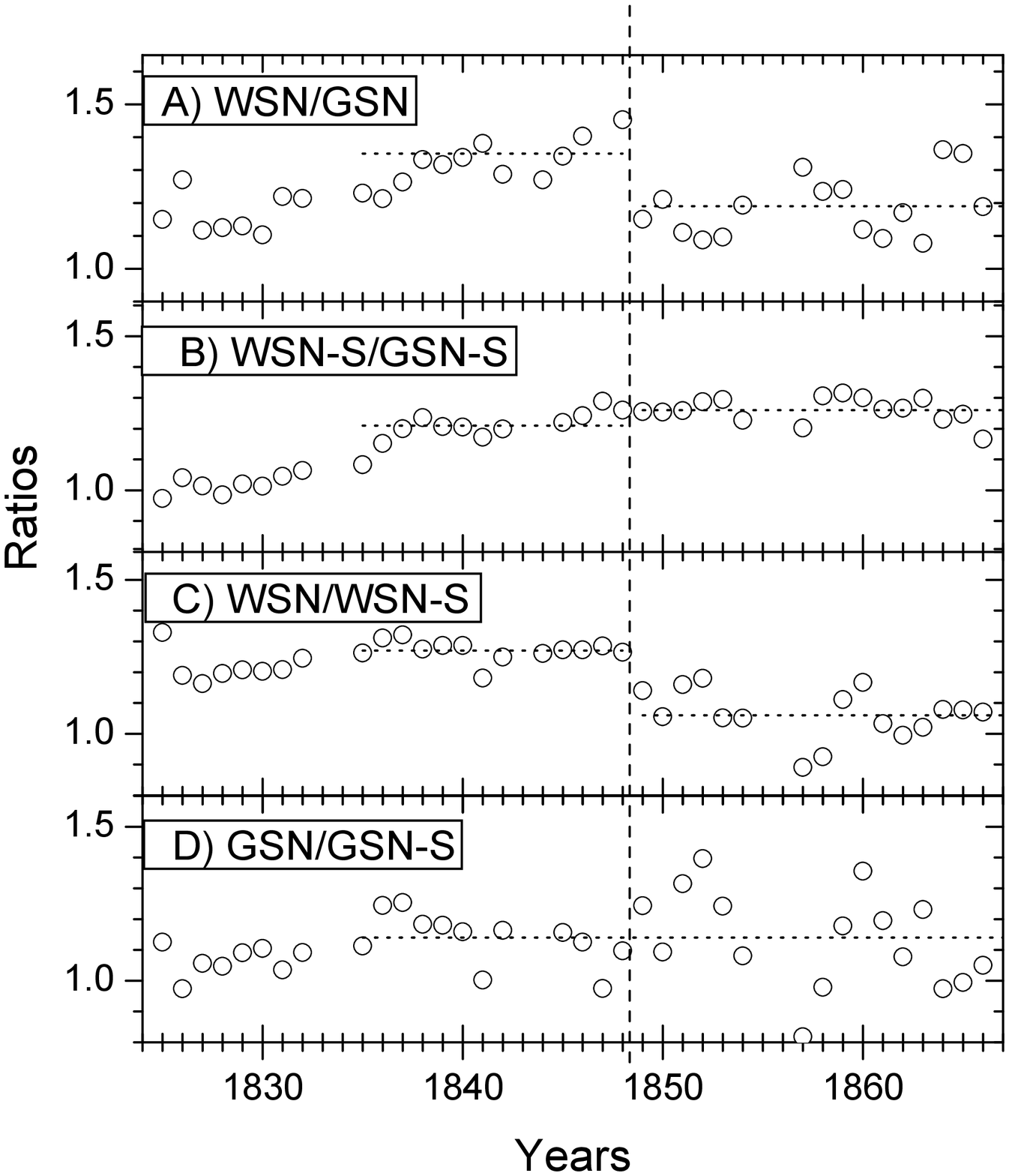} }
\caption{Ratios of the annual sunspot numbers:
a) the ratio of WSN to GSN;
b) the ratio of WSN-S to GSN-S;
c) the ratio of WSN to WSN-S;
d) the ratio of GSN to GSN-S.
 Vertical dashed line indicates the period when R. Wolf started his observations.
 Horizontal dotted lines depict the mean ratios before and after that date.
 Details of statistics are collected in Table~\ref{Tab:ratio}.}
\label{Fig:ratio}
\end{figure}

Here we compare the (annual values of) sunspot series as defined in Section~\ref{Sec:SN}, by analyzing their ratios in order to get rid of the scaling ambiguity.
In order to avoid large uncertainties related to division of small numbers, we omitted the years
 when the corresponding annual sunspot number in either series is smaller than 10.
The ratios are shown in Fig.~\ref{Fig:ratio}.

\begin{table*}
\caption{The mean ratios, along with its standard error,
 of different series of annual means (see Fig.~\ref{Fig:ratio}) before and after 1848,
 their difference $D$ and the significance level $s$ to reject the hypothesis that the
  ratio remains constant.
 \label{Tab:ratio}}
\begin{tabular}{lcccc}
Data & WSN/GSN & WSN-S/GSN-S & WSN/WSN-S & GSN/GSN-S\\
\hline
1835--1848 & $1.35\pm0.04$ & $1.21\pm0.02$ & $1.27\pm0.01$ & $1.14\pm0.03$\\
1849--1867 & $1.19\pm0.02$ & $1.26\pm0.01$ & $1.06\pm0.02$ & $1.14\pm0.04$\\
$D$ & $0.16\pm0.045$ & $0.05\pm0.022$ & $0.21\pm0.022$ & $0.0\pm0.05$\\
$s$ & $5\cdot 10^{-4}$ & $0.03$ & $< 10^{-7}$ & $0.96$\\
\hline
\end{tabular}
\end{table*}

\subsection{WSN vs. GSN}

First we compare the two official series,  WSN and GSN.
The ratio of the official Wolf sunspot numbers and the official group sunspot numbers is shown in Fig.~\ref{Fig:ratio}a.
A jump in the ratio is apparent around 1848, when the primary observer of the WSN series had changed from Schwabe to Wolf.
The jump from 1.35 to 1.19 is significant at the level of $5\cdot 10^{-4}$ (see Table~\ref{Tab:ratio}, column 2).
This indicates that the WSN and GSN series are mutually inconsistent but cannot verify the consistency of
 either series.

\subsection{WSN-S vs. GSN-S}

Next we analyze the ratio of WSN-S to GSN-S (Fig.~\ref{Fig:ratio}b and third column in Table~\ref{Tab:ratio}).
Both series are based solely on Schwabe's data and thus their ratio is expected to be stable.
One can see, however, that there is a small difference, likely related to the
 detection of small spots by Schwabe in the early part of the period, as discussed in Sect.~\ref{Sec:S} (Fig.~\ref{Fig:size}).
Otherwise the ratio is perfectly smooth with a small tendency to decrease somewhat around solar minima,
 probably because of the smaller size of spots.

\subsection{WSN vs. WSN-S}

The ratio between WSN and WSN-S is shown in Fig.~\ref{Fig:ratio}c and in the fourth column in Table~\ref{Tab:ratio}.
A sudden and highly significant jump is apparent at 1848--1849, exactly at the time when
 the primary observer of the WSN series was changed from Schwabe to Wolf.
This suggests that the WSN series suffers a calibration problem in the middle of the 19th century. The jump is highly significant (See Table~\ref{Tab:ratio}).

\subsection{GSN vs. GSN-S}

The ratio of GSN to GSN-S is shown in Fig.~\ref{Fig:ratio}d and in the fifth column in Table~\ref{Tab:ratio}.
It appears nearly constant throughout the entire period, confirming the homogeneity of the GSN series in the studied time interval.

\section{Discussion}

Let us now summarize the results of comparison of the four sunspot series, the two official ones and the two other based on Schwabe's observations.
The latter are perfectly homogeneous (at least after 1835 -- see Section~\ref{Sec:S}), being based on the
 systematic observations performed by an experienced observer using the same instrumentation
 and techniques.
This makes it possible to check the official series for homogeneity and internal consistency for the
 middle 19-th century.

The comparison shows that the GSN series is homogeneous around 1848, since the ratio between the official GSN series, which involves data from many observers, and
 the GSN-S series, based solely on Schwabe's data, is constant throughout the studied time interval.

\begin{figure}
\centering \resizebox{8cm}{!}{\includegraphics{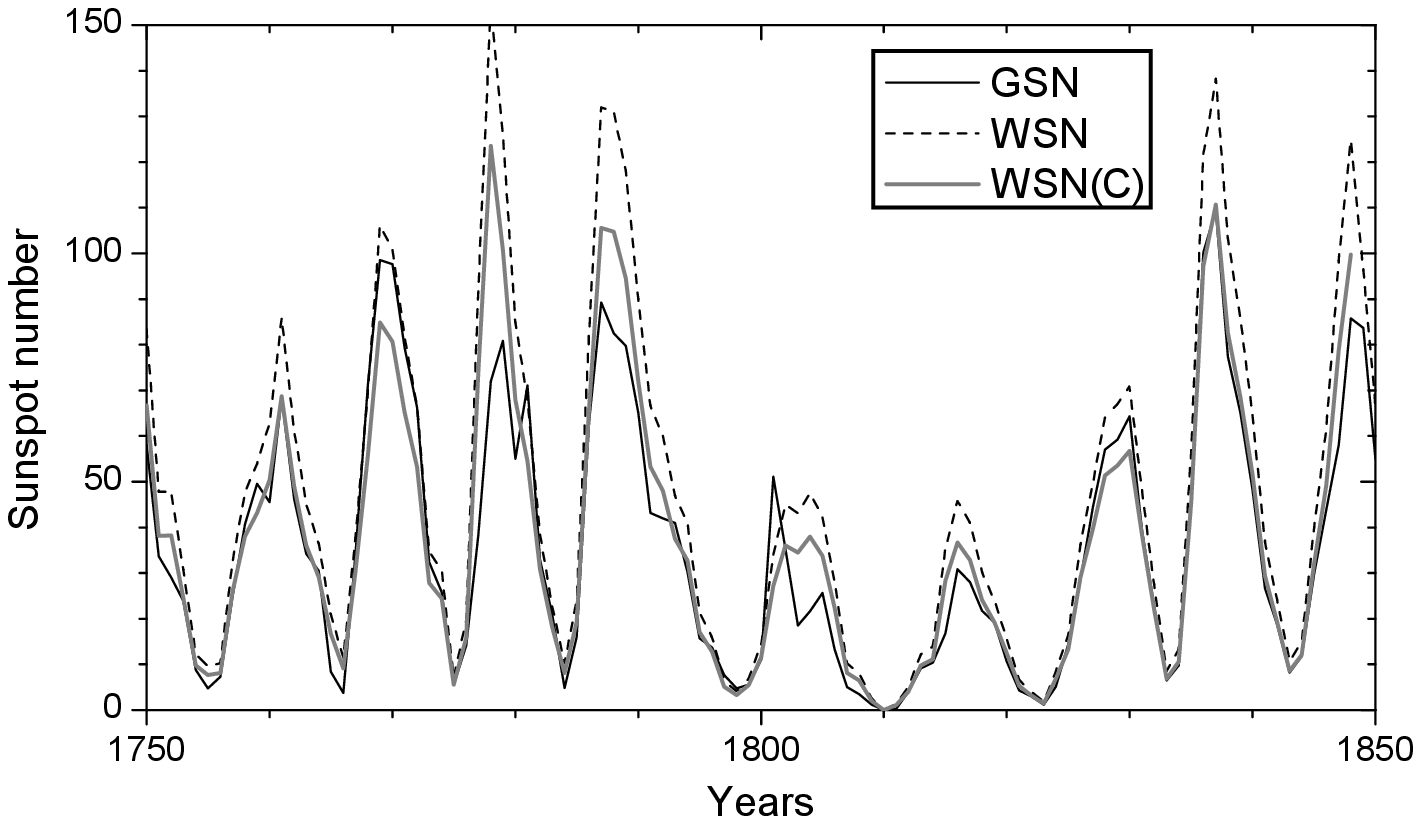} }
\caption{Annual sunspot numbers for the period 1750--1850 AD:
 The official GSN series (solid GSN line), the official WSN series (dashed WSN line)
 as well as by the 20\% reduced WSN series (grey WSN-C line).}
\label{Fig:corr}
\end{figure}

On the other hand, the WSN series appears inconsistent and experiences a significant decrease around 1848.
This decrease is observed relative to both the official GSN series (by 14\%) and WSN-S (by 20\%).
The latter suggests that the individual correction factor $k$ (see Eq.~\ref{Eq:WSN}) used by R. Wolf
 for Schwabe was inappropriate, and that it should be lowered by 20\% to make the WSN series homogeneous
 through the change of the primary observer in 1848.
This requires the corresponding reduction of the WSN values by 20\% for the period 1826--1848
 when Schwabe was the primary observer for WSN.
Moreover, since the ``calibration'' of the WSN series is consecutive in time using overlaps
 between observers, this leads to the 20\% reduction of the entire WSN series before 1848.
The WSN series corrected for this 20\% reduction is shown as the WSN-C series in Fig.~\ref{Fig:corr} along with
 the official WSN and GSN series.
One can see that the corrected WSN is in a much better agreement with the GSN than the official WSN series.
We note that most likely the main reason for this scatter in the ratios between these series is related to the difference between the individual observers and not to some intrinsic variation in the properties of sunspots or their distribution.
It should also be noted that the WSN might also need some correction for the period 1826-1835, as discussed earlier.

\section{Conclusions}

In this paper we have compared the official Wolf sunspot number and the group sunspot number series with the
 homogeneous multi-decadal record (1835--1867) of sunspot observations by Samuel Heinrich Schwabe, recently
 digitized by \citet{arlt13}.
We show that:
\begin{itemize}

\item
The GSN is homogeneous and consistent with Schwabe's data throughout the entire studied period.
\item
The WSN series suffers a significant $\approx 20$\% decrease around 1848 caused by the change of the
 primary observer from Schwabe to Wolf.
\item
The decrease reflects an inappropriate individual correction factor used for Schwabe in the WSN, and implies a major
 inhomogeneity in the WSN.
\item
The WSN needs to be corrected by decreasing its values by 20\% before 1848.
\item
Thus corrected WSN series is in good agreement with the GSN series.
\item
Before 1835 the WSN may be underestimated because Schwabe only considered large spots. A detailed study will be made later.
\end{itemize}

This study supports the earlier conclusions that the group sunspot number series \citep{hoyt98} is
 more consistent and homogeneous in the earlier part than the Wolf sunspot number series
  \citep[e.g.,][]{letfus99,letfus00,hathaway02,usoskin_SP03,usoskin13}.

\begin{acknowledgements}{The data for the official monthly Wolf sunspot number is from SIDC
 and the official monthly group sunspot number data from NOAA.
 RL acknowledges support from the V\"ais\"al\"a Foundation through the Finnish Academy of Science and Letters. }
\end{acknowledgements}


\end{document}